# The thermal gauge potentials in quantum transport

Zheng-Chuan Wang

The University of Chinese Academy of Sciences, P. O. Box 4588, Beijing 100049, China, wangzc@ucas.ac.cn

## Abstract

In 1964，Luttinger et al. proposed the thermal scalar and vector potential to study the thermal transport driven by the temperature gradient in terms of Kubo's linear response theory[1,2]. In this manuscript, we present another new thermal scalar and vector gauge potentials implemented by the quantum Boltzmann equation (QBE), which originates from the interaction of conduction electrons and phonons. To accomplish this task, we derive a temperature-dependent four-dimensional damping force by the Taylor series expansion on the self-energy of the QBE, which can be related to the thermal scalar and vector gauge potentials, especially the fourth component of the damping force, which is just a power corresponding to a new scalar potential. Based on the local equilibrium assumption, we solve the QBE order by order using the Fourier transformation method. The temperature-dependent damping force and other physical observables are exhibited in the figures, the higher of the temperature, the bigger of the damping force.

## I. Introduction

It is well known that the Yang-Mills gauge field[3] can be utilized to unify the electroweak and strong interactions by a standard model, which may be regarded as the connections in the principle bundle according to the fiber bundle theory[4]. Furthermore, the gravitational gauge field can also be described by the Cartan connection of the frame bundle in spacetime manifold[5]. In 1984, the gauge potential in the parameter space was pioneered by Berry in geometric phase theory[6], its integral along a path in the parameter space produced by the adiabatic variation of the quantum system will contribute a geometric phase, which is expounded as the holonomy concerning the fiber bundle in parameter space manifold[7], nowadays the Berry phase has played an important role in topological physics[8], i.e., topological insulator or quantum Hall effect et al. In the engineered electromagnetic system and quantum optical systems, the artificial gauge potential or the synthetic gauge potential is also proposed[9], which has no the kinetic energy term in the Lagrangian, only acts as a background field to drive the other fields and waves. There exists an emergent gauge field in the five-dimensional Kaluza-Klein gravity theory[10], too, which possesses the kinetic energy term in Lagrangian. This gauge field can unify the gravity theory and Maxwell's electromagnetic theory together.

The thermal scalar and vector potentials were also proposed in non-equilibrium thermodynamics, which is a gauge invariant to the physical observable. In 1964, Luttinger introduced a scalar "gravitational" potential to investigate the thermal transport driven by temperature gradient[1], the thermal transport coefficient can be obtained by Kubo's linear response theory with respect to this scalar potential. Since the scalar potential $\Psi$ is defined as $\vec{\nabla}\Psi = \frac{\vec{\nabla}T}{T}$, where $T(x)$ is the temperature distribution, the scalar "gravitational" potential is in fact a thermal potential that avoids the problem of adding the macroscopic temperature into the microscopic Hamiltonian. Until now, the thermal scalar potential had been widely utilized to explore the thermal transport of electrons, and Magnon et al.[11-15], Bauer et al. even employed it to investigate the thermally induced torque[16]. However, as mentioned in Ref.[11,13], the thermal transport coefficient expressed by the scalar potential will diverge when the temperature $T \to 0$. To overcome this unphysical divergence, Tatara proposed a thermal vector to replace Luttinger's scalar potential[2], which induced a diamagnetic current to overcome the divergence. In fact, as shown by wang[17] in the definition of scalar potential by Luttinger, the right-hand side term $\frac{\vec{\nabla}T}{T}$ has its curl, while the left-hand side term $\vec{\nabla}\Psi$ has no curl, so mathematically $\vec{\nabla}\Psi \neq \frac{\vec{\nabla}T}{T}$, we must introduce a vector potential to describe the curl of $\frac{\vec{\nabla}T}{T}$ instead using only the scalar potential. On the other hand, both the thermal scalar and vector potentials satisfy the gauge invariant property, they are the thermal gauge potentials.

Whatever the Luttinger's scalar potential or Tatara's vector potential is related to $\frac{\vec{\nabla}T}{T}$ which is in fact a macroscopic phenomenological thermal force. In Ref.[17], Wang presented other thermal scalar and vector gauge potentials microscopically based on QBE. They are related to the damping force, which originates from the scattering of the transport particles with the particles in the environment. They are also temperature dependent and keep the gauge invariant. Wang's thermal gauge potentials are expressed by the distribution functions of the transport and environmental particles. If we expand the distribution function by the electric field and the temperature gradient, the thermal current and electronic current can be connected with the thermal scalar and vector potential, which is conceded to Luttinger's work. Moreover, we can obtain the nonlinear corrections from the second- or higher- orders, which exceed Luttinger's linear response results.

Because Wang's microscopic thermal scalar and vector potentials originate from the scattering of transport particles, different scattering mechanisms will lead to different thermal gauge potentials. The scattering of the electron-impurity was investigated in Ref.[17], a temperature-dependent damping force can be derived based on QBE, which can be regarded as being contributed by the thermal gauge potentials. This three-dimensional force has its divergence and curl, so it is a damping force. If we further consider the spin freedom in the electron-impurity scattering and describe the transport procedure by the spinor Boltzmann equation, we can obtain a temperature-dependent spinor damping force, which can be expanded by the unit matrix and the three Pauli matrices[18], the coefficient of the unit matrix corresponds to the usual force, while the coefficients of the Pauli matrices correspond to the spinor part of the force, so the thermal gauge potentials concerning the spinor damping force will obtain their spinor indices, which is different from the thermal gauge potentials in Ref.[17]. In this manuscript, we explore the thermal gauge potential and the temperature-dependent damping force in the system with the electron-phonon interaction. We will show that a four-dimensional damping force will appear, which will induce a new thermal gauge scalar potential.

## II. Theoretical formalism

Consider the electronic transport under the external electric and magnetic fields in a conductor. There exists an electron-phonon interaction in this system, and the transport of conduction electrons can be described by the QBE given by Mahan:

$$i\{\frac{\partial}{\partial t} + \vec{v} \cdot \vec{\nabla} + e\vec{E} \cdot [(1 - \frac{\partial Re\Sigma^r}{\partial \omega})\vec{\nabla}_p + (\vec{v} + \vec{\nabla}_p Re\Sigma^r)\frac{\partial}{\partial \omega}]\}G^<(\vec{p},\omega,\vec{r},t) - ie\vec{E} \cdot [\frac{\partial \Sigma^<}{\partial \omega}\vec{\nabla}_p ReG^r - \frac{\partial ReG^r}{\partial \omega}\vec{\nabla}_p \Sigma^<] = [\Sigma^> G^< - \Sigma^< G^>] + i[Re\Sigma^r, G^<] + i[\Sigma^<, ReG^r],$$

(1)

where $G^>$ and $G^<$ are the greater and lesser Green functions, respectively, $G^< = if(\vec{p},\omega,\vec{r},t)$ and $G^> = if'(\vec{p},\omega,\vec{r},t)$, where $f(\vec{p},\omega,\vec{r},t)$ is the quantum Wigner distribution function, $f'(\vec{p},\omega,\vec{r},t)$ is the hole distribution function. $\Sigma^>$ and $\Sigma^<$ are the greater and lesser self-energies, $\Sigma^r$ and $G^r$ are the retarded self-energy and Green function, respectively. The bracket $[A,B]$, i.e. $[\Sigma^<, ReG^r]$, is defined as $\frac{\partial A}{\partial \vec{r}}\frac{\partial B}{\partial \vec{p}} - \frac{\partial A}{\partial \vec{p}}\frac{\partial B}{\partial \vec{r}}$. If we only consider the electron-phonon scattering in the system, the self-energy can be written as[19]

$$\Sigma^<(\vec{p},\omega) = \sum_q M_q^2 [(N_q + 1)G^<(\vec{p} + \vec{q}, \omega + \omega_q) + N_q G^<(\vec{p} + \vec{q}, \omega - \omega_q)]$$

(2)

and

$$\Sigma^>(\vec{p},\omega) = \sum_q M_q^2 [N_q G^>(\vec{p} + \vec{q}, \omega + \omega_q) + (N_q + 1)G^>(\vec{p} + \vec{q}, \omega - \omega_q)]$$

(3)

with $\Sigma^r(\vec{p},\omega) = \sum_q M_q^2 [\frac{N_q + 1 - n_{p+q}}{\omega - \varepsilon_{p+q} + \omega_q + i\delta} + \frac{N_q + n_{p+q}}{\omega + \varepsilon_{p+q} - \omega_q + i\delta}]$, where $N_q$ is the phonon occupation number, $n_{p+q}$ denotes the electron occupation number, and the matrix element $M_q$ depicts the strength of the electron-phonon interaction. Then, the first term in the right hand of Eq.(1) can be expressed as

$$-\sum_q M_q^2 [N_q f'(\vec{p} + \vec{q}, \omega + \omega_q) + (N_q + 1)f'(\vec{p} + \vec{q}, \omega - \omega_q)]f(\vec{p},\omega) + \sum_q M_q^2 [(N_q + 1)f(\vec{p} + \vec{q}, \omega + \omega_q) + N_q f(\vec{p} + \vec{q}, \omega - \omega_q)]f'(\vec{p},\omega),$$

(4)

where we have replaced the lesser and greater Green functions with their corresponding Wigner distribution functions. If the momentum transfer $\vec{q}$ caused by the electron-phonon scattering is small compared with the momentum $\vec{p}$ of electron, and the frequency of phonon $\omega_q$ is low, we can make a Taylor series expansion on the distribution function $f'(\vec{p}+\vec{q},\omega+\omega_q)$, $f'(\vec{p}+\vec{q},\omega-\omega_q)$, $f(\vec{p}+\vec{q},\omega+\omega_q)$ and $f(\vec{p}+\vec{q},\omega-\omega_q)$ around $\vec{p}$ and $\omega$ in Eq.(4), the zero order terms in the expansion will cancel with each other in the above two terms, only the first and second order terms are left, then expression (4) can be rewritten as

$$[\sum_q M_q^2 \omega_q]\frac{\partial f'}{\partial \omega}f - [\sum_q M_q^2 (2N_q+1)\vec{q}]\frac{\partial f'}{\partial \vec{p}}f - \frac{1}{2!}[\sum_q M_q^2 (2N_q+1)\omega_q^2]\frac{\partial^2 f}{\partial \omega^2}f + [\sum_q M_q^2 \vec{q}\omega_q]\frac{\partial^2 f'}{\partial \vec{p}\partial \omega}f - \frac{1}{2!}[\sum_q M_q^2 (2N_q+1)\vec{q}^2]\frac{\partial^2 f}{\partial \vec{p}^2}f + \frac{1}{2!}[\sum_q M_q^2 \vec{q}^2 \omega_q]\frac{\partial^3 f'}{\partial \vec{p}^2 \partial \omega}f + [\sum_q M_q^2 \omega_q]\frac{\partial f}{\partial \omega}f' + [\sum_q M_q^2 (2N_q+1)\vec{q}]\frac{\partial f}{\partial \vec{p}}f' + \frac{1}{2!}[\sum_q M_q^2 (2N_q+1)\omega_q^2]\frac{\partial^2 f}{\partial \omega^2}f' + [\sum_q M_q^2 \vec{q}\omega_q]\frac{\partial^2 f}{\partial \vec{p}\partial \omega}f' + \frac{1}{2!}[\sum_q M_q^2 (2N_q+1)\vec{q}^2]\frac{\partial^2 f}{\partial \vec{p}^2}f' + [\sum_q M_q^2 \vec{q}^2 \omega_q]\frac{\partial^3 f}{\partial \vec{p}^2 \partial \omega}f'.$$

(5)

If we rearrange Eq.(5) with the terms on the left-hand side of Eq.(1) and neglect the first order term $i[Re\Sigma^r, G^<] + i[\Sigma^<, ReG^r]$, then Eq.(1) can be rewritten as

$$\{\frac{\partial}{\partial t} + \vec{v}\cdot\vec{\nabla} + [(e\vec{E}+\vec{F}_{damp1})\vec{\nabla}_p + (e\vec{E}\cdot\vec{v}+\vec{F}_{damp2})\frac{\partial}{\partial \omega}]\}f(\vec{p},\omega,\vec{r},t) = f_{term}.$$

(6)

where $\vec{F}_{damp1} = -e\vec{E}\frac{\partial Re\Sigma^r}{\partial \omega} + [\sum_q M_q^2 (2N_q+1)\vec{q}]f' - [\sum_q M_q^2 (2N_q+1)]\frac{\partial ReG^r}{\partial \omega}$ and $\vec{F}_{damp2} = e\vec{E}\cdot\vec{\nabla}_p Re\Sigma^r + [\sum_q M_q^2 \omega_q]f'$. For brevity, we express $f_{term}$ in Appendix A.

Usually, the coefficient of $\vec{\nabla}_p f$ on the left-hand side of Eq.(1) corresponds to the force, so the term $\vec{F}_{damp1}$ is just the damping force as given by us in Ref.[17], while the coefficient $F_{damp2}$ of $\frac{\partial}{\partial \omega}f$ is the fourth component of the four-dimensional damping force $(\vec{F}_{damp1}, F_{damp2})$, which is a new term and in essence a power caused by the electron-phonon interaction. The four-dimensional damping force is formally similar to the four-dimensional force $(\vec{F}, \frac{i}{c}\vec{v}\cdot\vec{F})$ in electrodynamics, where its fourth component has a relationship with the other three components as $f_4 = \frac{i}{c}\vec{v}\cdot\vec{F}$, while in our four dimensional damping force, it is violated because they don't obey the Lorentz covariance.

In the scattering of electron-phonon, the phonon system can be regarded as a reservoir that remains in the local equilibrium thermodynamic state when the relaxation procedure of the phonon is faster than that of electrons; then, the phonon occupation number $N_q$ in Eq.(1) may be evaluated as $N_q = \frac{1}{\exp(\hbar\omega_q/kT(\vec{r}))-1}$, where k represents the Boltzmann constant and $T(\vec{r})$ depicts the temperature distribution of the reservoir. Thus, the temperature will appear in the phonon occupation number, and the terms $\vec{F}_{damp1}$ and $F_{damp2}$, which originate from the electron-phonon interaction, are temperature dependent. We refer to it as the damping force, as given in Ref.[20,21]. The temperature-dependent damping force is a thermal force, we expect it can be expressed by the thermal scalar and vector potentials.

We will show that both the scalar and the vector potentials contribute to the damping force because it is a dissipative force with divergence and curl, and not a conservative force. For the damping force $\vec{F}_{damp1}$, if we let the scalar potential satisfies

$$\nabla^2 \varphi = \vec{\nabla} \cdot \vec{F}_{damp1}, \qquad (7)$$

and the vector potential obeys

$$-\frac{\partial(\vec{\nabla}\times\vec{A})}{\partial t} = \vec{\nabla} \times \vec{F}_{damp1}, \qquad (8)$$

the results show that both the thermal scalar potential $\varphi$ and vector potential $\vec{A}$ depend on the temperature. They are determined by the Wigner distribution function in the temperature-dependent damping force and are thus thermal potentials. Consequently, from Eq.(7) and (8), we obtain the following:

$$-\frac{\partial \vec{A}}{\partial t} - \vec{\nabla}\varphi = \vec{F}_{damp1}. \qquad (9)$$

When we perform a gauge transformation $\varphi \to \varphi - \dot{\chi}$, $\vec{A} \to \vec{A} + \vec{\nabla}\chi$, where $\chi$ is a scalar function, Eq.(9) can keep the gauge invariant, $\varphi$ and $\vec{A}$ are the thermal gauge potentials.

For the damping force $F_{damp2}$, we introduce another scalar potential $\psi$ and let

$$\frac{\partial \psi}{\partial t} = F_{damp2}, \qquad (10)$$

then $\psi = \int F_{damp2} dt$, it is a new scalar potential for the four-dimensional damping force, differs from the scalar potential $\varphi$ in Eq.(7). The four-dimensional damping force is temperature dependent, and the scalar potentials $\varphi$, $\psi$ and vector $\vec{A}$ are also temperature dependent. If we choose a position $\vec{r}$ dependent temperature $T(\vec{r}) = T_0 + b\vec{r}$ in the system, where b is the temperature gradient, $T_0$ is the temperature at the boundary of the system, then the thermal scalar and vector potentials are concerning with the temperature gradient according to Eqs.(9) and (10), which is consistent with the thermal potentials proposed by Luttinger[1] and Tatara[2]. Our results can be reduced to Luttinger's in the first-order approximation, as shown in Ref.[17].

Next, we investigate the influence of the quantum correction terms on the right-hand side of Eq.(1). Similar to the zero-order terms discussed above, the first order terms $i[Re\Sigma^r, G^<] + i[\Sigma^<, ReG^r]$ can also be

expanded by a Taylor series expansion, which yields

$$-\frac{\partial ReG^r}{\partial \vec{p}}\{[\sum_q M_q^2 (2N_q + 1)]\frac{\partial f}{\partial \vec{r}} + [\sum_q M_q^2 \omega_q]\frac{\partial^2 f}{\partial \vec{r}\partial \omega} + [\sum_q M_q^2 (2N_q + 1)\vec{q}]\frac{\partial^2 f}{\partial \vec{r}\partial \vec{p}} + \frac{1}{2!}[\sum_q M_q^2 (2N_q + 1)\omega_q^2]\frac{\partial^3 f'}{\partial \omega^2 \partial \vec{r}} + [\sum_q M_q^2 \vec{q}\omega_q]\frac{\partial^3 f}{\partial \vec{r}\partial \vec{p}\partial \omega} + \frac{1}{2!}[\sum_q M_q^2 (2N_q + 1)\vec{q}^2]\frac{\partial^3 f}{\partial \vec{p}^2 \partial \vec{r}} + \frac{1}{2!}[\sum_q M_q^2 \vec{q}^2 \omega_q]\frac{\partial^4 f}{\partial \vec{p}^2 \partial \omega \partial \vec{r}}\} + \frac{\partial ReG^r}{\partial \vec{r}}\{[\sum_q M_q^2 (2N_q + 1)]\frac{\partial f}{\partial \vec{p}} + [\sum_q M_q^2 \omega_q]\frac{\partial^2 f}{\partial \vec{p}\partial \omega} + [\sum_q M_q^2 (2N_q + 1)\vec{q}]\frac{\partial^2 f}{\partial \vec{p}^2} + \frac{1}{2!}[\sum_q M_q^2 (2N_q + 1)\omega_q^2]\frac{\partial^3 f}{\partial \omega^2 \partial \vec{p}} + [\sum_q M_q^2 \vec{q}\omega_q]\frac{\partial^3 f}{\partial \vec{p}^2 \partial \omega} + \frac{1}{2!}[\sum_q M_q^2 (2N_q + 1)\vec{q}^2]\frac{\partial^3 f}{\partial \vec{p}^3} + \frac{1}{2!}[\sum_q M_q^2 \vec{q}^2 \omega_q]\frac{\partial^4 f}{\partial \vec{p}^3 \partial \omega}, \quad (11)$$

then QBE containing the quantum correction terms can be rewritten as follows:

$$\{\frac{\partial}{\partial t} + (\vec{v} + \vec{v}_{anor}) \cdot \vec{\nabla} + [(e\vec{E} + \vec{F}'_{damp1})\vec{\nabla}_p + (e\vec{E} \cdot \vec{v} + F'_{damp2})\frac{\partial}{\partial \omega}]\}f(\vec{p}, \omega, \vec{r}, t) = f'_{term},$$

(12)

where $\vec{v}_{anor} = -\frac{\partial Re\Sigma^r}{\partial \vec{p}} + [\sum_q M_q^2 (2N_q + 1)]\frac{\partial ReG^r}{\partial \vec{p}}$ is the anomalous velocity contributed by the quantum correction terms, which is similar to the anomalous velocity in anomalous Hall effect[22]. $\vec{F}'_{damp1} = -e\vec{E}\frac{\partial Re\Sigma^r}{\partial \omega} + [\sum_q M_q^2 (2N_q + 1)\vec{q}]f' - [\sum_q M_q^2 (2N_q + 1)]\frac{\partial ReG^r}{\partial \omega} + [\sum_q M_q^2 (2N_q + 1)]\frac{\partial ReG^r}{\partial \vec{r}}$ is the damping force and $F'_{damp2} = e\vec{E} \cdot \vec{\nabla}_p Re\Sigma^r + [\sum_q M_q^2 \omega_q]f' - [\sum_q M_q^2 (N_q + 1)]e\vec{E} \cdot \vec{\nabla}_p ReG^r$ is fourth component of the four-dimensional damping force ($\vec{F}'_{damp1}, F'_{damp2}$). Thus the quantum correction terms can also contribute to the damping force. For brevity, we provide the expression of $f'_{term}$ in Appendix B.

### III. Numerical results

It is difficult to find a solution for Eq.(6). For simplicity, we consider only the electronic transport through a one-dimensional conductor. Eq.(6) can be solved using the Fourier transform method. Based on the local equilibrium assumption, the distribution function of the conduction electrons can be expanded around the local equilibrium distribution function $f_0(p, x)$ as:

$$f(p, x, t) = f_0(p, x) + f_1 + f_2 + ..., \quad (13)$$

where $f_1$ and $f_2$ are the first- and second-order deviation away from the local equilibrium distribution. As a starting point of the iteration procedure, we first approximate the term $\vec{\nabla}_p f$ and $\frac{\partial f}{\partial \omega}$ by $\vec{\nabla}_p f_0$ and $\frac{\partial f_0}{\partial \omega}$, and omit the higher-order terms in the equation, then Eq.(6) become to

$$\frac{\partial f_1}{\partial t} + \vec{v} \cdot \vec{\nabla} f_1 = -\frac{\partial f_0}{\partial t} - \vec{v} \cdot \vec{\nabla} f_0 - (e\vec{E} + \vec{F}_{damp1})\vec{\nabla}_p f_0 - (e\vec{E} \cdot \vec{v} + F_{damp2})\frac{\partial f_0}{\partial \omega} + f_{term0},$$

(14)

where $f_{term0}$ denotes the distribution function $f$ in $f_{term}$ is approximated by $f_0$. Eq.(14) can be solved by the Fourier transformation method. If we obtain the first- order correction $f_1$, then we can express Eq.(14) as

$$\frac{\partial f_2}{\partial t} + \vec{v} \cdot \vec{\nabla} f_2 = -\frac{\partial (f_0+f_1)}{\partial t} - \vec{v} \cdot \vec{\nabla}(f_0 + f_1) -$$
$$(e\vec{E} + \vec{F}_{damp1})\vec{\nabla}_p (f_0 + f_1) - (e\vec{E} \cdot \vec{v} +$$
$$\vec{F}_{damp2})\frac{\partial (f_0+f_1)}{\partial \omega} + f_{term1}, \quad (15)$$

where $f_{term1}$ represents the distribution function $f$ in $f_{term}$ is approximated by $(f_0 + f_1)$. Step by step, we can calculate the higher order terms in Eq.(13) order by order. After we obtain the distribution function, we may evaluate the physical observable through their statistical expressions.

The physical observable, e.g., the charge current density and thermal current density et al., can be calculated by the above approximated solution, and we will demonstrate it numerically. There exists temperature gradient $b = 2K/nm$ in this system, and the position-dependent temperature is adopted as $T(x) = T_0 + bx$, where $T_0$ is the temperature at the boundary. In our calculation, the electric field is chosen as $E = -1.0 \times 10^4 V \cdot m^{-1}$. In Fig. 1, we plot the damping force $\vec{F}_{damp1}$ as a function of position at different temperatures 100K, 200K, 300K. The magnitude of damping force is comparable with the external electric field force, so we cannot neglect it. This damping force originates from the scattering terms in the QBE;

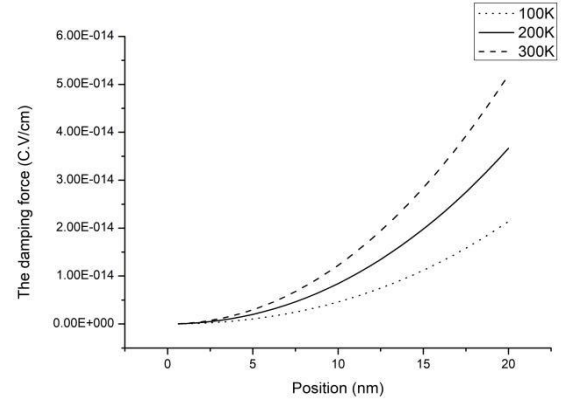

Fig.1 The damping force vs position at temperature $T_0 = $ 100K, 200K, 300K.

In our system, it comes from the scattering of the electron-phonon. Since we only study the one-dimensional transport, when we apply the external electric field along the x-axis, the vector part of the four-dimensional damping force only has the x-component left.

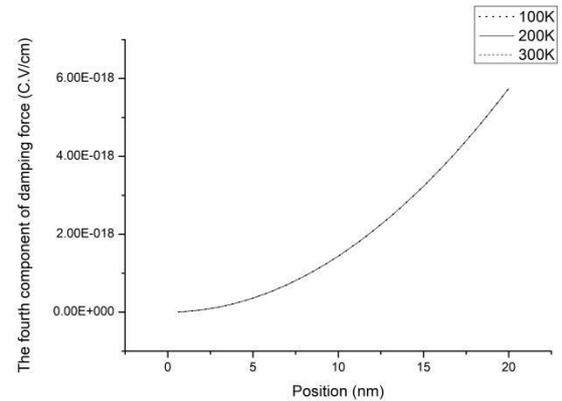

Fig.2 The fourth component of damping force vs position at temperature $T_0 = $ 100K, 200K, 300K.

As pointed in Eq.(7) and (8), the damping force have its divergence and curl, so it is position dependent. The higher of the temperature, the bigger of the damping force. The fourth-component of the four-dimensional damping force is shown in Fig. 2, it increases with position, which is in essence a power, so it is a scalar mathematically, we can see that it varies with temperature slightly. This damping component also comes from the electron-phonon scattering. The charge current density vs. position is shown in Fig. 3, and it varies with the position. It seems that the current density is different at the left and right boundaries, but the current is still governed by the charge continuity equation. The variation in the current density is caused by the electron-phonon scattering, which is the origin of the electric resistance in this system. The thermal density as a function of position is shown in Fig. 4 at the temperatures 100K, 200 K, 300K, they change with temperature obviously, the higher of the temperature, the bigger of the magnitude of the thermal current density. In our calculation the thermal current is only carried by the electrons, so it is analogous to the electronic current.

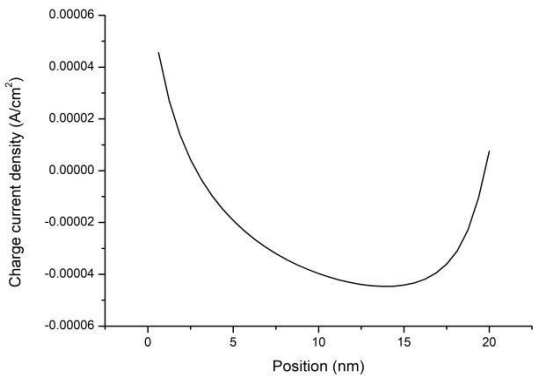

Fig.3 The charge current density vs position at temperature $T_0$ =200K.

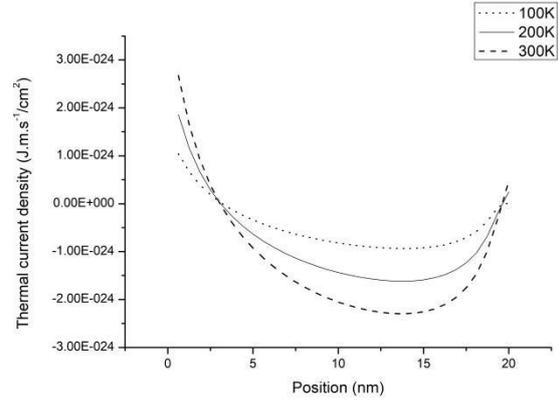

Fig.4 The thermal current density vs position at temperatures $T_0 = $ 100K, 200K, 300K.

### IV. Summary and discussions

In this manuscript, we deal with deriving a four-dimensional damping force based on the QBE developed by Mahan, which can be expressed by the thermal scalar and vector gauge potentials. The fourth component of the damping force and the corresponding scalar potential are the new results. We also demonstrate the influence of quantum corrections, which contribute not only to the four-dimensional damping force but also to the velocity terms, resulting in an anomalous velocity term similar to Luttinger's anomalous Hall effect. After adopting the local equilibrium assumption, we solve the QBE order by order by the Fourier transformation method, and the physical observables are shown in Figs.1-4, the higher of the temperature, the bigger of the damping force.

It is worth emphasizing that our thermal scalar and vector potentials are different from those of Luttinger and Tatara. They originate from the scattering of conduction electrons with the particles in the reservoir, such as electron-impurity scattering and electron-phonon scattering et al., while the latter come from the temperature gradient and is a macroscopic phenomenological thermal force, they must exist in the system with temperature gradient. Our thermal gauge potentials can even exist in the system without a temperature gradient, i.e., the system driven by the external electric field, because the temperature in our thermal gauge potential is introduced by the reservoir, which interacts with the conduction electrons. Since the reservoir remains in the local equilibrium and can be described by a temperature that will appear in the damping force and thermal gauge potentials finally. Certainly, our thermal gauge potential can also exist in the system with a temperature gradient. In this case, we can adopt the local equilibrium assumption in the reservoir and introduce the temperature gradient into the distribution function of the particles, as shown above. Our thermal gauge potentials can be expressed by these distribution functions of the conduction particles and the particles in the reservoir, so our thermal gauge potentials can be naturally related to the temperature gradient, they are different from Luttinger's thermal scalar potential which is defined as $\vec{\nabla}\Psi = \frac{\vec{\nabla}T}{T}$, its relationship with temperature and temperature gradient is direct, somehow it is a replacement of temperature gradient in order to be put in the Hamiltonian.

### Acknowledgments

This study is supported by the National Key R&D Program of China (Grant No. 2022YFA1402703).
### Acknowledgments

This study is supported by the National Key R&D Program of China (Grant No. 2022YFA1402703).

### Data Availability Statement

Data sets generated during the current study are available from the corresponding author on reasonable request.

# Appendix A

$$f_{term} = \left\{ -\left[\sum_q M_q^2 \omega_q\right]\frac{\partial f'}{\partial \omega} + \left[\sum_q M_q^2 (2N_q+1)\vec{q}\right]\frac{\partial f'}{\partial \omega}\right.$$

$$+ \frac{1}{2!}\left[\sum_q M_q^2(2N_q+1)\omega_q^2\right]\frac{\partial^2 f'}{\partial \omega^2} - \left[\sum_q M_q^2 \vec{q}\omega_q\right]\frac{\partial^2 f'}{\partial \vec{p}\partial \omega}$$

$$+ \frac{1}{2!}\left[\sum_q M_q^2(2N_q+1)\vec{q}^2\right]\frac{\partial^2 f'}{\partial \vec{p}^2} - \frac{1}{2!}\left[\sum_q M_q^2 \vec{q}^2 \omega_q\right]\frac{\partial^3 f'}{\partial \vec{p}^2 \partial \omega}\right\}f$$

$$- \frac{1}{2!}[\sum_q M_q^2(2N_q+1)\omega_q^2]f'\frac{\partial^2 f}{\partial \omega^2} + i[\sum_q M_q^2 \vec{q}\omega_q]f\frac{\partial^2 f}{\partial \vec{p}\partial \omega}$$

$$- i\{-i\frac{1}{2!}\left[\sum_q M_q^2(2N_q+1)\vec{q}^2\right]f' + [\sum_q M_q^2(2N_q+1)\vec{q}]\}\frac{\partial^2 f}{\partial \vec{p}^2}$$

$$- i\{-i[\sum_q M_q^2 \vec{q}^2 \omega_q]f' + [\sum_q M_q^2 \vec{q}\omega_q]\}\frac{\partial^3 f}{\partial \vec{p}^2 \partial \omega} - i[\sum_q M_q^2(2N_q$$

$$+ 1)]\frac{\partial^2 f}{\partial \vec{p}\partial \vec{r}} - i\frac{1}{2!}[\sum_q M_q^2(2N_q+1)\omega_q^2]\frac{\partial^3 f}{\partial \vec{r}\partial \omega^2}$$

$$- i[\sum_q M_q^2 \vec{q}\omega_q]\frac{\partial^3 f}{\partial \vec{p}\partial \vec{r}\partial \omega} - i\frac{1}{2!}[\sum_q M_q^2(2N_q+1)\vec{q}^2]\frac{\partial^3 f}{\partial \vec{p}^2 \partial \vec{r}}$$

$$- i\frac{1}{2!}[\sum_q M_q^2 \vec{q}^2 \omega_q]\frac{\partial^4 f}{\partial \vec{p}^2 \partial \omega \partial \vec{r}} - i\frac{1}{2!}[\sum_q M_q^2(2N_q+1)\omega_q^2]\frac{\partial^3 f}{\partial \vec{p}\partial \omega^2}$$

$$- i\frac{1}{2!}[\sum_q M_q^2(2N_q+1)\vec{q}^2]\frac{\partial^3 f}{\partial \vec{p}^2} - i\frac{1}{2!}[\sum_q M_q^2 \vec{q}^2 \omega_q]\frac{\partial^4 f}{\partial \vec{p}^3 \partial \omega}$$

# Appendix B

$$f'_{term} = \left\{ -\left[\sum_q M_q^2 \omega_q\right]\frac{\partial f'}{\partial \omega} + \left[\sum_q M_q^2 (2N_q + 1)q\right]\frac{\partial f'}{\partial \omega}\right.$$

$$+ \frac{1}{2!}\left[\sum_q M_q^2 (2N_q + 1)\omega_q^2\right]\frac{\partial^2 f'}{\partial \omega^2} - \left[\sum_q M_q^2 \vec{q}\omega_q\right]\frac{\partial^2 f'}{\partial \vec{p}\partial \omega}$$

$$\left. + \frac{1}{2!}\left[\sum_q M_q^2(2N_q + 1)\vec{q}^2\right]\frac{\partial^2 f'}{\partial \vec{p}^2} - \frac{1}{2!}\left[\sum_q M_q^2 \vec{q}^2\omega_q\right]\frac{\partial^3 f'}{\partial \vec{p}^2\partial \omega}\right\}f$$

$$- i\left\{-i\frac{1}{2!}\left[\sum_q M_q^2 (2N_q + 1)\omega_q^2\right]f' + \left[\sum_q M_q^2 \omega_q\right]e\vec{E}\cdot\vec{\nabla}_p ReG^r\right\}\frac{\partial^2 f}{\partial \omega^2}$$

$$- i\left\{-i\left[\sum_q M_q^2 \vec{q}\omega_q\right]f' + ie\vec{E}\cdot\left[\sum_q M_q^2 \omega_q\right]\frac{\partial ReG^r}{\partial \vec{r}}\right.$$

$$\left. - i\left[\sum_q M_q^2 (2N_q + 1)\vec{q}\right]e\vec{E}\cdot\vec{\nabla}_p ReG^r\right\}\frac{\partial^2 f}{\partial \vec{p}\partial \omega} - i\left\{-i\frac{1}{2!}[\sum_q M_q^2 (2N_q\right.$$

$$\left. + 1)\vec{q}^2]f' + [\sum_q M_q^2 (2N_q + 1)\vec{q}] + i[\sum_q M_q^2 (2N_q + 1)\vec{q}]e\vec{E}\frac{\partial ReG^r}{\partial \vec{r}}\right\}\frac{\partial^2 f}{\partial \vec{p}^2}$$

$$- i\{-i[\sum_q M_q^2 \vec{q}^2\omega_q]f' + [\sum_q M_q^2 \vec{q}\omega_q] - \frac{1}{2!}[\sum_q M_q^2 (2N_q + 1)\vec{q}^2]e\vec{E}$$

$$\cdot \vec{\nabla}_p ReG^r] + ie\vec{E}\cdot[\sum_q M_q^2 \vec{q}\omega_q]\frac{\partial ReG^r}{\partial \vec{r}}\}\frac{\partial^3 f}{\partial \vec{p}^2\partial \omega}$$

$$+ [\sum_q M_q^2 \omega_q]\vec{\nabla}_p ReG^r \frac{\partial^2 f}{\partial \vec{r}\partial \omega} - i[\sum_q M_q^2 (2N_q + 1)]\frac{\partial^2 f}{\partial \vec{p}\partial \vec{r}}$$

$$- i\frac{1}{2!}[\sum_q M_q^2 (2N_q + 1)\omega_q^2]\frac{\partial^3 f}{\partial \vec{r}\partial \omega^2} - i[\sum_q M_q^2 \vec{q}\omega_q]\frac{\partial^3 f}{\partial \vec{p}\partial \vec{r}\partial \omega}$$

$$- i\frac{1}{2!}[\sum_q M_q^2 (2N_q + 1)\vec{q}^2]\frac{\partial^3 f}{\partial \vec{p}^2\partial \vec{r}} - i\frac{1}{2!}[\sum_q M_q^2 \vec{q}^2\omega_q]\frac{\partial^4 f}{\partial \vec{p}^2\partial \omega\partial \vec{r}}$$

$$-i\{\frac{1}{2!}[\sum_q M_q^2(2N_q+1)\omega_q^2] - i[\sum_q M_q^2 q\omega_q]e\vec{E}\cdot\vec{\nabla}_p ReG^r] - \frac{1}{2!}[\sum_q M_q^2(2N_q+1)\omega_q^2]\frac{\partial ReG^r}{\partial \vec{r}}\}\frac{\partial^3 f}{\partial \vec{p}\partial \omega^2} - i\{\frac{1}{2!}\left[\sum_q M_q^2(2N_q+1)\vec{q}^2\right]$$

$$- ie\vec{E}\frac{1}{2!}\left[\sum_q M_q^2(2N_q+1)\vec{q}^2\frac{\partial ReG^r}{\partial \vec{r}}\right\}\frac{\partial^3 f}{\partial \vec{p}^3}$$

$$- i\left\{\frac{1}{2!}\left[\sum_q M_q^2 \vec{q}^2 \omega_q\right] - ie\vec{E}\frac{1}{2!}\left[\sum_q M_q^2 \vec{q}^2 \frac{\partial ReG^r}{\partial \vec{r}}\right\}\frac{\partial^4 f}{\partial \vec{p}^3 \partial \omega}$$

$$- \frac{1}{2!}\left[\sum_q M_q^2(2N_q+1)\omega_q^2\right]e\vec{E}\cdot\vec{\nabla}_p ReG^r\right]\frac{\partial^3 f}{\partial \omega^3} - \frac{1}{2!}[\sum_q M_q^2 \vec{q}^2 \omega_q]e\vec{E}\cdot\vec{\nabla}_p ReG^r]\frac{\partial^4 f}{\partial \vec{p}^2 \partial \omega^2}$$